\documentclass[prb,twocolumn,showpacs]{revtex4}
\usepackage{bm}
\usepackage{graphicx}
\usepackage{amsmath}
\usepackage{eufrak}
\usepackage{color}
\usepackage{ulem}
\newcommand{\nix}[1]{}
\begin{document}

\title{Resonant circular photogalvanic effect in GaN/AlGaN heterojunctions}

\author{B.~Wittmann,$^1$ L.\,E.~Golub,$^2$  S.\,N.~Danilov,$^1$ J.~Karch,$^1$
C.~Reitmaier,$^1$  Z.\,D.~Kvon,$^3$ N.\,Q.~Vinh,$^4$ A.\,F.\,G.~van~der~Meer,$^4$ B.~Murdin,$^5$
and S.\,D.~Ganichev$^{1}\footnote{e-mail:
sergey.ganichev@physik.uni-regensburg.de}$}
\affiliation{$^1$ Terahertz Center, University of Regensburg, 93040
Regensburg, Germany}
\affiliation{$^2$ A.F.~Ioffe Physico-Technical Institute, Russian
Academy of Sciences, 194021 St.~Petersburg, Russia}
\affiliation{$^3$ Institute of Semiconductor Physics, Russian Academy
of Sciences, 630090 Novosibirsk, Russia}

\affiliation{$^4$ FOM Institute for Plasma Physics ``Rijnhuizen'', P.O.
Box 1207, NL-3430 BE Nieuwegein, The Netherlands}

\affiliation{$^5$ University of Surrey, Guildford, GU2 7XH, UK}

\pacs{ 73.21.Fg, 78.67.De, 73.63.Hs, 72.25.Dc}

\begin{abstract}
The resonant circular photogalvanic effect is
observed in wurtzite (0001)-oriented
GaN low-dimensional structures excited by infrared radiation.
The current is induced by angular momentum transfer of photons to the
photoexcited electrons at resonant inter-subband optical transitions in a GaN/AlGaN heterojunction.
The signal reverses upon the reversal of the radiation helicity or, at fixed helicity,
when the propagation direction of the photons is reversed. Making use of the
tunability of the free-electron laser FELIX we demonstrate that
the current direction changes by sweeping the photon energy through
the intersubband resonance condition, in agreement with theoretical considerations.
\end{abstract}

\date{\today}

\maketitle

\section{Introduction}

Wide bandgap GaN has been extensively investigated for  applications as
blue and ultraviolet light sources~\cite{Nakamurabook} as well as for  high temperature
and high power electronic devices.~\cite{GaN1,GaN2,GaN3} The commercial fabrication of
blue and green LEDs has led to well established technological
procedures of epitaxial GaN preparation and sparked a great research
activity on the properties of heterostructures based
on GaN and its alloys with AlN and InN. Most recently two-dimensional
GaN attracted growing attention as a potentially interesting material system
for semiconductor spintronics since, doped with manganese,
it is expected to become ferromagnetic with a Curie-temperature above
room temperature,~\cite{Dietl2000} gadolinium doped it may offer an opportunity
for fabricating magnetic semiconductors,\cite{7gado,7gadobis,8gado,9gado} and
since GaN-based structures show long spin relaxation times~\cite{Beschoten01} and
considerable Rashba spin splitting~\cite{Rashba} due to strong built-in electric
fields.
First indications of substantial spin-orbit splitting
came from the observation of the circular photogalvanic effect
(CPGE)~\cite{bookIvchenko,bookGanichevPrettl} in GaN
heterojunctions at Drude absorption of THz radiation~\cite{APL2005} and
was then confirmed by
magneto-transport measurements
yielding a value for the Rashba splitting of about 0.3~meV at the Fermi
wavevector.~\cite{ThillosenAPL2006,SchmultPRB2006,TangAPL2006}
Making use of interband absorption, the investigations of the CPGE were extended to GaN quantum wells
as well as to low dimensional structures under uniaxial strain, confirming the
Rashba character of the spin splitting.~\cite{HeAPL2007,TangApl2007,ChoPRB2007}

In this paper we report on the observation of the resonant
CPGE due to inter-subband transitions and present the phenomenological
theory as well as the microscopic model of this phenomenon.
We demonstrate that variation of the photon energy in the vicinity
of the resonance results in a change of
sign of the photocurrent. This proves the dominant contribution  to the total current from the
asymmetry in momentum distribution of carriers excited in optical
transitions. We analyze spin-dependent as well
as spin-independent mechanisms giving rise to a
resonant photocurrent and demonstrate that, in spite of the weak spin-orbit
interaction, the resonant CPGE in GaN is
mostly caused by the spin-dependent mechanism.

\section{Samples and experimental methods}

%

\begin{figure}[t]
\includegraphics[width=0.6\linewidth]{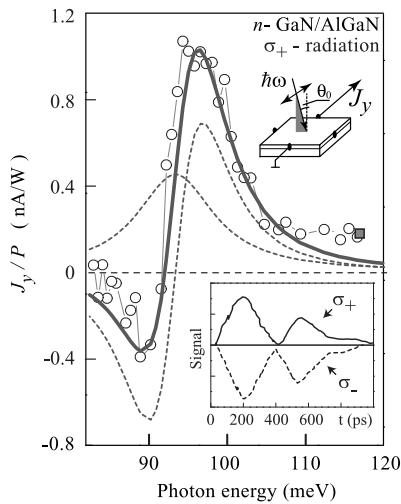}
\caption{Spectral dependence of the transversal photocurrent $J_y$ measured
at room temperature at oblique incidence ($\theta_0 = 15^\circ$)
for right-handed circularly polarized radiation.
The data are obtained by the free electron laser FELIX (dots) and TEA CO$_2$ laser (square).
The radiation power used for these measurements was about $35$~kW for the CO$_2$ laser and
about 100~kW  for FELIX. The magnitude of the current measured at FELIX is fit at this plot to that
obtained with the  CO$_2$ laser, assuming that the current depends linearly  on the radiation power at used power level.
The solid line represents a fit by the sum of asymmetrical and symmetrical
contributions to Eq.~\protect(\ref{j}).
The  inset in the bottom corner shows the temporal structure of the current in response
to the radiation of FELIX. The inset in the {upper} corner demonstrates
the experimental geometry.
}
 \label{resonatCPGEinver}
\end{figure}

Experiments were carried out on GaN/Al$_{0.3}$Ga$_{0.7}$N
heterojunctions grown by MOCVD on C(0001)-plane sapphire
substrates (for details of growth see Ref.~\onlinecite{APL2005}). The thickness
of the AlGaN layers was varied between 30~nm and
100~nm. An undoped 33~nm thick GaN buffer layer grown
under a pressure of 40~Pa at temperature 550$^{\circ}$C is
followed by an undoped GaN layer ($\sim $ 2.5~$\mu $m) grown under
40~Pa at 1025$^{\circ}$C; the undoped Al$_{0.3}$Ga$_{0.7}$N
barrier was grown under 6.7~Pa at 1035$^{\circ}$C. The
mobility and  density in the two-dimensional electron
gas measured at room temperature are $\mu \approx 1200$~cm$^{2}$/Vs
and ${N_e} \approx 10^{13}$~cm$^{-2}$, respectively. To measure
the photocurrent two  pairs of contacts are centered at opposite
sample edges with the connecting lines along the axes $x \parallel [11\bar{2}0]$
and $y \parallel [\bar{1}100]$, see inset in the top of Fig.~\ref{resonatCPGEinver}.
In order to excite resonantly transitions between the size quantized subbands $e1$ and $e2$ and to obtain a
measurable photocurrent it was necessary to have a tunable high
power radiation source for which we used the free electron laser
``FELIX'' at FOM-Rijnhuizen in the Netherlands operating in the
spectral range between 8~$\mu$m and 15~$\mu$m.~\cite{Knippels99p1578}
The output pulses of light
from FELIX were chosen to be about 3~ps long, separated by 40~ns, in a
train (or ``macropulse'') of duration of 7~$\mu$s. The macropulses
had a repetition rate of 5~Hz. The linearly polarized light from FELIX was converted into left handed ($\sigma_-$) and right handed
($\sigma_+$) circularly polarized radiation by means of a Fresnel rhomb.
A rotation of the optical axis  of the Fresnel rhomb's plate by an angle $\varphi$
with respect to the laser radiation polarization plane results in a variation of
the radiation helicity as  $P_{\rm circ} = \sin{2 \varphi}$. Here the angle $\varphi = 0$ corresponds to the setting where the
position of the polarizer optical axis coincides with the incoming laser polarization.
Radiation is applied at oblique incidence described by an
angle of incidence $\theta_0$ varying from $-15^\circ$ to
+15$^\circ$, see the top inset to Fig.~\ref{resonatCPGEinver}.

\section{Experimental results and Phenomenological Description}

\begin{figure}[t]
\includegraphics[width=0.8\linewidth]{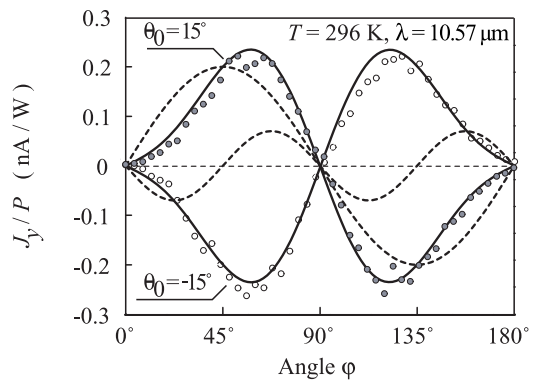}
\caption{Photocurrent as a function of the angle $\varphi$ measured at oblique incidence ($\theta_0 = \pm 15^\circ$) in
the transverse geometry. The data are obtained in close circuit configuration
applying TEA CO$_2$ laser operating   at a wavelength $\lambda$~=~10.57~$\mu$m
and a power $P \approx 35$~kW.
Solid lines are fits of the photocurrent to Eq.~\protect(\ref{phi-comp}) at $\gamma/\chi =1.4$.
The two contributions to Eq.~\protect(\ref{phi-comp}) are shown by dashed lines for $\theta_0 =  15^\circ$.
}
 \label{phiRT10p6mkm}
\end{figure}

On illumination of the low-dimensional structures by circularly polarized
radiation at room temperature at oblique incidence in ($xz$)-plane we observe a current signal  in
the $y$ direction perpendicular to the plane of incidence, Fig.~\ref{resonatCPGEinver}.
The current reverses its direction by switching the sign
of the radiation helicity. Signals in response to the left handed and right handed
polarized radiation are shown in the lower inset of Fig.~\ref{resonatCPGEinver}. Using short
3~ps pulses of  FELIX we observed that the response time was
determined by the time resolution of our set-up which therefore sets an upper limit to the response time.
This fast response is typical for photogalvanics, where the signal decay time is
expected to be of the order of the momentum relaxation time~\cite{bookIvchenko,bookGanichevPrettl,SturmanFridkin}
being in our samples at room temperature typically about $0.1$~ps.
Figure~\ref{phiRT10p6mkm} demonstrates the dependences of the photocurrent on the  angle $\varphi$
for two angles of incidence $\theta_0 = \pm 15^\circ$. The photocurrent signals generated in the
unbiased devices were measured via an amplifier with a response
time of the order of 1 $\mu$s, i.e. averaged over the macropulse.
The current closely follows
the radiation helicity $P_{\rm circ} = \sin{2 \varphi}$.
The signal proportional to the helicity is only observed under oblique incidence.
The current vanishes for normal incidence and changes its polarity when the incidence angle changes
its sign, see Fig.~\ref{phiRT10p6mkm}. The photocurrent in the layer flows always perpendicularly
to the direction of the incident
light propagation and its magnitude does not change by rotating the sample around the growth axis.

Investigating the spectral dependence of the photocurrent we observe that
for both, left and right handed circular
polarizations, the current changes sign at a photon energy $\hbar \omega \approx 93$~meV,
see  Fig.~\ref{resonatCPGEinver}.
Figure~\ref{resonatCPGEinver} also shows that the spectral behaviour of the current
can be well described by the sum of the derivative of the
Lorentzian-like absorption spectrum and an additional contribution proportional to the absorption spectrum itself.

A phenomenological analysis for the C$_{3v}$ point group symmetry relevant to the structures under study
shows that the $\varphi$-dependence of the transverse photogalvanic current density $\bm j$
under excitation at oblique incidence in ($xz$)-plane
is given by~\cite{APL2005,PRB2008}
\begin{eqnarray} \label{phi-comp}
j_y({\varphi}) = E_0^2 t_p t_s \sin{\theta} \: \left(\gamma \sin{2\varphi} -{\chi \over 2} \sin{4\varphi} \right) \,.
\end{eqnarray}
Here $E_0$
is the amplitude of the
electromagnetic wave,
$\theta$ is the refraction angle related to the incidence angle $\theta_0$
by $\sin{\theta} = \sin{\theta_0}/n_\omega$, where $n_\omega$ is the refractive index,
and $t_s$ and $t_p$ are the Fresnel amplitude  transmission coefficients from vacuum to the structure for
the $s$-  and $p$-polarized light, respectively.~\cite{Born_Wolf} The coefficient $\gamma$
is a component of the second-rank pseudotensor
$\bm \gamma$ which describes the helicity dependent current, comprising the CPGE
and the optically induced spin-galvanic effect,\cite{Nature2002} and $\chi$ is a component of the third-rank
tensor $\bm \chi$ describing the linear photogalvanic effect.\cite{bookIvchenko,bookGanichevPrettl,PRB2008}
In systems of $C_{3v}$ symmetry the
tensor $\bm \gamma$ has one linearly-independent component,
namely $\gamma_{xy} = - \gamma_{yx} \equiv \gamma$. Thus, the CPGE current
flows always perpendicularly to the plane of incidence. The components of the tensor $\bm \chi$
are given by  $\chi \equiv \chi_{xxz} = \chi_{yyz}$, where $z \parallel [0001]$ is the $C_3$ axis.
We take into account that the second linearly-independent component of $\bm \chi$ in $C_{3v}$ systems is much smaller than $\chi$, so that the corresponding photocurrent is negligible at oblique incidence.\cite{PRB2008}

The fits of the experiment to Eq.~\eqref{phi-comp} at $\gamma/\chi =1.4$ are shown
in Fig.~\ref{phiRT10p6mkm} for  $\theta_0 = \pm 15^\circ$
by solid lines demonstrating a good agreement  with the experimental data.
Figure~\ref{phiRT10p6mkm} shows that the dominant contribution to the photocurrent is due to the
helicity dependent current given by the  first term
on the right-hand side of the Eq.~\eqref{phi-comp}. The fact that the magnitude of the photocurrent
does not change by rotating the sample around the growth axis is in agreement with the fact that
in structures of C$_{3v}$ symmetry the
tensor $\bm \gamma$ has one linearly-independent component.

\section{Microscopic models and Discussion}

Two microscopic mechanisms are known to give rise to
helicity-dependent currents exhibiting sign reversal for photon energies
matching  the energy separation
between size-quantized subbands.
The first is caused by an asymmetry of
the momentum distribution of  carriers excited by direct optical transitions
in the system with the Rashba splitting of energy
subbands.~\cite{bookIvchenko,bookGanichevPrettl} Spectral inversion at resonance
is a characteristic feature of this CPGE mechanism
and has also been observed in GaAs QWs.~\cite{invCPGE2003,pureSGE2003}
The second mechanism is of orbital nature and spin-independent. It is caused by
the quantum mechanical interference of various pathways in Drude absorption.~\cite{ST_CPGE_orbital}
This effect has most recently been demonstrated in Si-MOSFET inversion
layers, where spin dependent mechanisms
are absent due to vanishingly small spin-orbit interaction in silicon.~\cite{MOSFET2008}
The spectral inversion in orbital mechanism is caused by a difference of the virtual states in
the excited subband needed for the pathway with
direct virtual optical transitions.
 Below we consider both mechanisms and
compare their contributions to the total photocurrent.

\begin{figure}[t]
\includegraphics[width=0.95\linewidth]{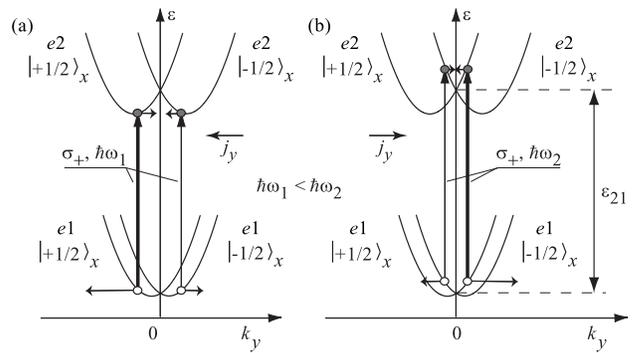}
\caption{Microscopic picture describing the origin of the CPGE and
its spectral sign inversion in C$_{3v}$ point group samples. (a)
Excitation at oblique incidence with $\sigma_+$ radiation of
$\hbar \omega$ less than the subband energy  separation
$\varepsilon_{21}$ induces direct spin-conserving transitions
(vertical arrows).
The rates of these
transitions  are different as illustrated by the different
thickness of the arrows (reversing the angle of incidence mirrors
the  thickness of the arrows). This leads to a photocurrent due to an
asymmetrical distribution of carriers in {\boldmath $k$}-space if
the splitting of the $e1$ and $e2$ subbands are not equal. (b)
Increasing of the photon energy shifts more intensive transitions to
the right and less intensive to the left, resulting in a change of
the photocurrent
sign.
}
 \label{model}
\end{figure}

The spin-dependent mechanism of the CPGE caused by direct inter-subband transitions
at oblique incidence is illustrated in Fig.~\ref{model} for $\sigma_+$ radiation.
In C$_{3v}$ symmetry the $\alpha\sigma_{x} k_y$ term in the Rashba
Hamiltonian splits the subbands in the $k_y$ direction into two
spin branches with the spin projection $\pm1/2$ oriented along $x$.
Due to selection rules, like in (001)-grown GaAs QWs of C$_{2v}$ symmetry,
the absorption of circularly polarized radiation is
spin-conserving.~\cite{invCPGE2003} It turns out, however,  that under oblique
excitation with circularly polarized light
the rates of inter-subband transitions are different for electrons
with the spin oriented parallel and antiparallel to the in-plane
direction of light propagation.~\cite{invCPGE2003}
This is depicted in
Fig.~\ref{model} by vertical arrows of different
thickness. In systems with ${\mbox{\boldmath$k$}}$-linear spin
splitting such processes lead to an asymmetrical distribution of
carriers in ${\mbox{\boldmath$k$}}$-space, i.e. to an electrical
current. The inversion of photon helicity driven current is a direct
consequence of {\boldmath $k$}-linear terms in the band structure
of subbands together with energy and momentum conservation as well
as optical selection rules for direct optical transitions between
size quantized subbands.
At  photon energy $\hbar\omega_1 < \varepsilon_{21}$ of right
circularly polarized radiation the most intense optical transition occurs at
negative $k_y$ resulting in a current $j_y$
shown by an arrow in Fig.~\ref{model}~(a). Here $\varepsilon_{21}$ is the
intersubband energy separation. Increase of
the photon energy shifts the transition toward positive $k_y$
and reverses the direction of the current, see Fig.~\ref{model}~(b).
In the frame of this model the
sign reversal of the current takes place  at the photon energy
$\hbar \omega = \varepsilon_{21}$ corresponding to optical transitions from
the band minima.
The CPGE current at direct inter-subband transitions is given by~\cite{invCPGE2003}
\begin{eqnarray}\label{j}
    j_{\rm spin} &=& \Lambda  (\alpha_1-\alpha_2 ) {e  \over \hbar} {I P_{\rm circ} \over \hbar\omega}
    \\
    && \times \left[ (\tau_p^{(1)} - \tau_p^{(2)}) \bar{E} {d \eta_{21}(\hbar\omega) \over d \hbar\omega}
    + \tau_p^{(2)}
    \eta_{21}(\hbar\omega)\right]. \nonumber
\end{eqnarray}
Here $\alpha_{1,2}$ are the Rashba constants for electrons in the
first and second subbands, the factor $\Lambda$ describes monopolar spin orientation of
carriers under resonant transitions, $\tau_p^{(1)}$ and $\tau_p^{(2)}$
are the momentum relaxation times in the initial and final state of optical transition,
$\bar{E}$ is the average kinetic energy of carriers,
$\eta_{21}(\hbar\omega)$ is the intersubband absorbance which in the model of an
infinitely-deep rectangular quantum well is given by~\cite{bookIvchenko}
\[
    \eta_{21}(\hbar\omega) = {512 \over 27 \pi} {N_e e^2\hbar\over c m n_\omega}{\Gamma \over (\varepsilon_{21}-\hbar \omega)^2 + \Gamma^2},
\]
where $m$ is the electron effective mass and $\Gamma$ is the peak width.
The 2DEG in GaN structures is almost degenerate even at room temperature due to the large Fermi energy:
$\bar{E} \approx 100~\mbox{meV} > k_{\rm B}T$. Therefore, in contrast to GaAs based structures, the
optical phonon emission by photoexcited electrons in the $e2$ subband
is suppressed (also due to the high frequency of the optical phonon in GaN).
As a result, both initial and final states have momentum relaxation times of the same order, so that
both symmetrical [$\propto \eta_{21}(\hbar\omega)$] and asymmetrical
[$\propto d\eta_{21}/d(\hbar\omega)$] terms give comparable contributions to the CPGE.
This is due to the fact that GaN based low dimensional structures
have a large energy of optical phonons and typically a rather large Fermi energy compared to the GaAs QWs.

The orbital mechanism of the CPGE is caused by Drude like absorption which is
usually rather small at infrared frequencies
used in the experiment. However, its contribution to the total photocurrent may
be comparable with the spin-dependent mechanism and, therefore, should be analyzed.
Following Refs.~\onlinecite{ST_CPGE_orbital,MOSFET2008} the orbital mechanism of the CPGE is described by
\begin{equation}\label{j_Drude}
    j_{\rm orb} = \xi z_{21}
    {4\pi\kappa e^3 N_e \over \omega c m n_\omega}
    {\varepsilon_{21}\over \varepsilon_{21}^2 -(\hbar \omega)^2}
    I P_{\rm circ},
\end{equation}
where $\xi$ is the factor of scattering asymmetry, and $\kappa \gtrsim 1$.~\cite{ST_CPGE_orbital}
This equation shows that the CPGE due to the orbital mechanism also changes the photocurrent sign
for photon energies matching the energy separation.

To compare spin-dependent and orbital contributions to the CPGE we take magnitudes of the photocurrent at
the wings of the absorption contour ($|\hbar \omega -  \varepsilon_{21}| = \Gamma$).
Equations~\eqref{j} and~\eqref{j_Drude} give an estimate for the ratio of the contributions as
\[
j_{\rm spin}/j_{\rm orb} \sim {\Lambda \over \xi} {(\alpha_1-\alpha_2) \omega \tau_p^{(1)}  \over \Gamma z_{21}} {\bar{E}\over \varepsilon_{21}}.
\]
The factor $\Lambda$ describing monopolar spin orientation of
carriers under resonant transitions with account for both crystal
($\Delta_{cr}$) and spin-orbit ($\Delta$) splittings of the valence band
provided $\Delta \ll \Delta_{cr} \ll E_g$ ($E_g$ is the fundamental energy gap) is given by
$$\Lambda = {\varepsilon_{21} \: \Delta \over 3 E_g^2}.$$
It can be estimated for investigated GaN based structures as $\Lambda \sim 0.03$.
This substantially smaller value of $\Lambda$ in GaN structures compared to GaAs QWs is caused by a
small spin-orbit interaction in the nitrogen atom, yielding $\Delta/E_g \approx 10^{-3}$ for GaN.
Taking $\varepsilon_{21}=100$~meV and $\Gamma=6$~meV from the photocurrent spectrum,
$z_{21} = 10$~\AA,\cite{Litvinov2003}
$|\alpha_1-\alpha_2| \sim \alpha_1 \approx 10^{-10}$~eV~cm,~\cite{ThillosenAPL2006,SchmultPRB2006}
and $\xi=10^{-2}$ as in Ref.~\onlinecite{MOSFET2008}, we obtain that the spin-dependent
contribution to CPGE caused by resonant direct transitions
is about five times larger than the CPGE due to the orbital mechanism caused by quantum
interference in Drude absorption.

This conclusion is also supported by the shape of the CPGE spectrum.
Indeed, Fig.~\ref{resonatCPGEinver} shows that the
shape of the spectrum is substantially asymmetric.
The orbital mechanism described by Eq.~\eqref{j_Drude}
yields only a slight asymmetry while for the spin-dependent mechanism in GaN structures we obtain that
the photocurrent is a superposition of comparable symmetric [$\propto \eta_{21}(\hbar\omega)$] and asymmetric
[$\propto d\eta_{21}/d(\hbar\omega)$] parts, Eq.~(\ref{j}). Figure~\ref{resonatCPGEinver} shows that
Eq.~(\ref{j}) describes well the whole spectral behaviour. We note that, when
considering the spin dependent contribution to the helicity dependent photocurrent,
one should also take into account a possible admixture of the spin-galvanic effect which is
proportional to the radiation absorbance and $P_{\rm circ}$.
The interplay between CPGE and
the spin-galvanic effect caused by resonant inter-subband transitions has been
reported for GaAs QWs.~\cite{pureSGE2003}


To summarize, we demonstrate that in GaN low dimensional structures resonant
intersubband transitions result in a circular photogalvanic
effect with a dominant contribution by the spin dependent mechanism. The specific feature
of the resonant CPGE in GaN heterojunctions is that the symmetric and asymmetric components of the photocurrent
have comparable strengths.

\acknowledgments
We thank E.L. Ivchenko, V.V.~Bel'kov and S.A.~Tarasenko for their permanent interest in this activity
and the many discussions that helped to clarify the problem under study. The financial support
from the DFG and RFBR is gratefully acknowledged.
Work of L.E.G. is also supported by ``Dynasty'' Foundation --- ICFPM and President
grant for young scientists. The high quality GaN samples were kindly provided by
Hyun-Ick Cho and Jung-Hee Lee from Kyungpook National University, Korea.

\end{document}